\documentclass[twocolumn,secnumarabic,amssymb, nobibnotes, aps, prd]{revtex4}
\usepackage{graphicx}

\begin{document}
\title{Aharonov-Bohm effects in nanostructures}
\author{V.L. Gurtovoi, A.V.  Nikulov, and V.A. Tulin}
\affiliation{Institute of Microelectronics Technology and High Purity Materials, Russian Academy of Sciences, 142432 Chernogolovka, Moscow District, RUSSIA.} 
\begin{abstract} Measurements of the Little-Parks oscillations at measuring current much lower than the persistent current give unambiguous evidence of the dc current flowing against the force of the dc electric field because of the Aharonov-Bohm effect. This result can assume that an additional force is needed for description of the Aharonov-Bohm effect observed in semiconductor, normal metal and superconductor nanostructures in contrast to the experimental result obtained recently for the case of the two-slit interference experiment.
 \end{abstract}

\maketitle

\narrowtext

\section*{Introduction}

Y. Aharonov and D. Bohm have shown in 1959 [1] that according to the universally recognized quantum formalism magnetic flux can act on quantum-mechanical state of charged particles even if the flux is enclosed and the particles do not cross any magnetic field lines [2]. The Aharonov-Bohm effect proposed [1] and observed [3] first for the two-slit interference experiment becomes apparent in numerous mesoscopic quantum phenomena observed in semiconductor and metal nanostructures [4-12]. This effect has fundamental impotence and is considered [13] as one of the most remarkable attainment in the centenarian history of quantum physics. The influence of the magnetic vector potential $A$ on the phase $\varphi $ of the wave function $\Psi = |\Psi |\exp{i\varphi }$ results directly from the universally recognized interpretation of the phase gradient $\bigtriangledown \varphi $ as a value proportional to the canonical momentum $\hbar \bigtriangledown \varphi = p = mv + qA$ of a particle with the mass $m$ and the charge $q$. But the non-local force-free quantum momentum $p = \hbar \bigtriangledown \varphi $ transfer implied in the Aharonov-Bohm effect provokes debates [14-16] which is bucked up in the last years [17,18]. Because of the phase shift $\Delta \varphi = q\Phi /\hbar $ observed in this phenomenon it is seem that magnetic forces can act on charged particles such as electrons - even though the particles do not cross any magnetic field lines [17]. Some authors assume such forces [16]. The direct experimental test [18] has give unambiguous evidence that the Ahranov-Bohm effect in the case of the two-slit interference experiment can not be connected with a force which could shift the phase $\Delta \varphi = q\Phi /\hbar $. 

One of the consequence of the Aharonov-Bohm effect in nanostructures is the persistent current observed in semiconductor, normal metal and superconductor loops [2,7-9,19-21]. The persistent current is observed in agreement with the theories [22-24] obtained in the limits of the universally recognized quantum formalism. But its observations in the loops with non-zero resistance call in question the force balance as well as in the case of the Ahranov-Bohm phase shift in the two-slit interference experiment [1-3,14-16]. Both these problems can not be considered as solved. The experimental evidence of the absence of time delays associated with forces of the magnitude needed to explain the Ahranov-Bohm phase shift [18] rather makes the problem more urgent than solves it. In the case of the Aharonov-Bohm effect in nanostructures the problem can be made more manifest because of the possibility of a dc potential difference $V$ on loop halves with non-zero resistance $R > 0$. The observation of the persistent current $I_{p} \neq 0$ in this case could mean that it can flow against the force of electric field $E = - \bigtriangledown V$ if it can be considered as a direct circular equilibrium current. The progress of nanotechnology allows to make such investigation which together with the investigation of the two-slit interference experiment [17,18] can give new information on paradoxical nature of the Aharonov-Bohm effect. Semiconductor and metal nanostructures could be used for this investigation. We use superconductor nanostructure consisting system of aluminum rings with radius $r \approx 1 \ \mu m$.  

\section {Experimental evidence of the persistent current as a direct circular equilibrium current.}
First of all one should note that the persistent current is observed as a direct circular equilibrium current in many experiments [7,8,19-21,25]. The periodical change of sign and magnitude of the magnetization $M = SI_{p}$ [7,8,19] and the dc voltage $V_{p} \propto I_{p}$ [20,21,25] with the period of magnetic field $H_{0}$ corresponding to the flux quantum $\Phi _{0} = 2\pi \hbar /q$ inside the loop with the area $S$ gives unambiguous evidence that the persistent current has clockwise or anti-clockwise direction depending on magnetic field magnitude $H$. The amplitude $I_{p,A}$ of the $I_{p}(H) \approx I_{p,A}2(n - \Phi /\Phi _{0})$ oscillations, increasing with loop perimeter $l$ decreasing [8], does not exceed $I_{p,A} = 1 \ nA$ in semiconductor and normal metal loop with the perimeter $l \approx 4 \ \mu m$ [19]. In superconductor loop the $I_{p,A}$ value is much higher even above superconducting transition $T > T_{c}$ [7], where $R > 0$ and the persistent current is observed because of the thermal fluctuations. Magnetization measurements of aluminum rings with section $s = 6000 \ nm^{2}$ (110-nm-wide 60-nm-thick) have revealed at $T = T_{c}$ the oscillations $I_{p}(H)$ with the amplitude $I_{p,A} \approx  700 \ nA$ at radius $r = l/2\pi \approx 0.5 \ \mu m$, $I_{p,A} \approx  200 \ nA$ at $r  \approx 1 \ \mu m$ [7] corresponding to the theoretical prediction [24]. Our measurements of the critical temperature shift $\Delta T_{c}$ induced by the external current and the persistent current have collaborated these results in order of value, Fig.1. 

\begin{figure}[]
\includegraphics{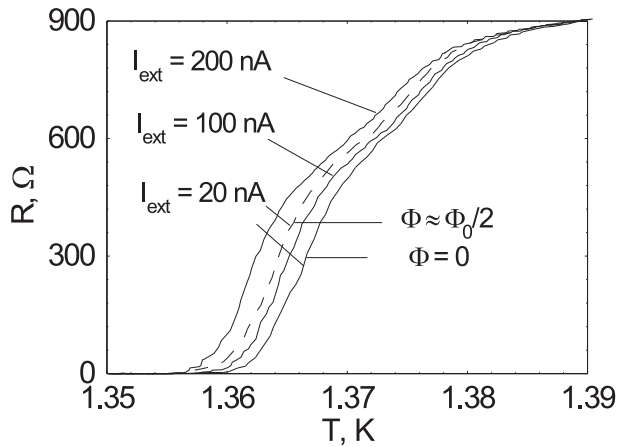}
\caption{\label{fig:epsart} The resistive transition $R(T) = V/I_{ext}$ of a system of 110 aluminium rings connected in series with radius $r \approx  1 \  \mu m$ and half-ring sections $s_{n} = 4000 \ nm^{2}$  (200-nm-wide 20-nm-thick) and $s_{n} = 8000 \ nm^{2}$  (400-nm-wide 20-nm-thick)  measured at different values of the measuring current $I_{ext} = 20 \ nA; 100 \ nA; 200 \ nA$ and the persistent current $I_{p} = 0$ at $\Phi = 0$, $I_{p} \approx 100 \ nA$ at $\Phi = \Phi _{0}/2$. The $R(T)$ shift $-\Delta T_{c} \approx 0.0025 \ K$  induced by the persistent current $I_{p}$ with maximum value observed at  $\Phi = \Phi _{0}/2$ is large than the one $-\Delta T_{c} \approx 0.0015 \ K$ induced by the external current $I_{ext} = 100 \ nA$ and smaller than $-\Delta T_{c} \approx 0.004 \ K$  induced by $I_{ext} = 200 \ nA$. }
\end{figure}

\begin{figure}
\includegraphics{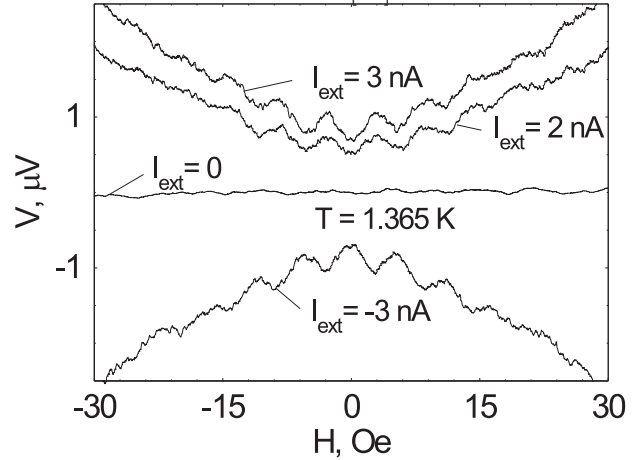}
\caption{\label{fig:epsart} The Little-Parks oscillations of the potential difference $V = RI_{ext}$ measured on a system of 110 aluminium rings connected in series at low measuring current $I_{ext} = -3 \ nA; 0 \ nA; 2 \ nA; 3 \ nA$ and the temperature $T \approx  1.365 \ K$ corresponding to the lower part of the resistive transition $R(T)$. The resistance $R = V/I_{ext}$ oscillations in magnetic field $R(H)$ do not depend on sign and magnitude of the measuring current $I_{ext}$ and is consequence $\Delta R(H) \propto I_{p}^{2}(H)$ of the persistent current oscillations $I_{p}(H)$ with the amplitude $I_{p,A} \approx  100 \ nA$. The period of the oscillations $H_{0} = 5.2 \ Oe$ corresponds to the flux quantum inside the ring with the area $S = \pi r^{2} = 4 \ \mu m^{2}$. }
\end{figure}

\section{Measurements of the Little-Parks oscillations at low measuring current}
In order to observe the Little-Parks oscillations $R(H)$ [26] the ring resistance is found as the relation $R = V/I_{ext}$ of the potential difference $V$ measured on ring halves to the measuring current $I_{ext}$ [25]. The dc electric filed $E = -\bigtriangledown V$ measured on both ring halves is directed from left to right or from right to left depending on the $I_{ext}$ direction. The circular persistent current (and the total current $I = I_{p} - I_{ext}/2$ at $I_{ext} < 2I_{p}$ ) is directed against the force of the dc electric field $E = -\bigtriangledown V$ in one of the ring halves since the $I_{p} \neq 0$ is observed at a magnetic flux $\Phi \neq n\Phi _{0}$ constant in time $d\Phi /dt = 0$. In order to observe the Little-Parks oscillations at low measuring current $I_{ext} \ll I_{p}$ we used a system with great number of rings connected in series [25]. We could not observe the $R(H)$ oscillations at $I_{ext} < 50 \ nA$ in [25] because of a noise. Additional shielding has allowed us the measure $R(H)$ at $I_{ext} = 2 \ nA \ll  I_{p,A} \approx  100 \ nA$, Fig.2. These results give unambiguous evidence of the dc current $I = I_{p} - I_{ext}/2 \approx  I_{p}$  flowing against the force of the dc electric field $E = -\bigtriangledown V  \approx  - RI_{ext}/\pi r$ because of the Aharonov-Bohm effect. The Aharonov-Bohm effect in this case can not be explained without an additional force in contrast to the result [18] obtained for the case on the two-slit interference experiment.   

\section*{Acknowledgement}
This work has been supported by a grant "Possible applications of new mesoscopic quantum effects for making of element basis of quantum computer, nanoelectronics and micro-system technic" of the Fundamental Research Program of ITCS department of RAS, the  Russian Foundation of Basic Research grant 08-02-99042-r-ofi and a grant of the Program "Quantum Nanostructures" of the Presidium of RAS.


\begin{thebibliography}{99}

\bibitem{Tulin01} Y. Aharonov and D. Bohm, {\em Phys. Rev.} {\bf 115}, 485 (1959).

\bibitem{Tulin02} S. Olariu and I. I. Popescu, {\em Rev. Mod. Phys.} {\bf 57}, 339 (1985)

\bibitem{Tulin03} R. G. Chambers, {\em Phys. Rev. Lett.} {\bf 5}, 3 (1960).

\bibitem{Tulin04} Doing-In Chang et al., {\em Nature Phys.} {\bf 4}, 205 (2008).  

\bibitem{Tulin05} F. Loder et al., {\em Nature Phys.} {\bf 4}, 112 (2008). 

\bibitem{Tulin06} R. Matsunaga et al., {\em Phys. Rev. Lett.} {\bf 101}, 147404 (2008)

\bibitem{Tulin07} N. C. Koshnick et al., {\em Science} {\bf 318}, 1440 (2007).

\bibitem{Tulin08} N. A. J. M. Kleemans et al., {\em Phys. Rev. Lett.} {\bf 99}, 146808 (2007)

\bibitem{Tulin09} V. M. Fomin et al., {\em Phys. Rev. B} {\bf 76}, 235320 (2007).

\bibitem{Tulin10} V. L. Campo Jr et al., {\em in Proceedings of 15th International Symposium "NANOSTRUCTURES: Physics and Technology"} St Petersburg: Ioffe Institute, 2007 p. 236

\bibitem{Tulin11} O.~A.~Tkachenko et al., {\em in Proceedings of 14th International Symposium "NANOSTRUCTURES: Physics and Technology"} St Petersburg: Ioffe Institute, 2006, p. 250.

\bibitem{Tulin12} D.~V.~Nomokonov et al., {\em in Proceedings of 13th International Symposium "NANOSTRUCTURES: Physics and Technology"} St Petersburg: Ioffe Institute, 2005, p. 197; O.~A.~Tkachenko et al., {\em idid}, p. 205

\bibitem{Tulin13} D. Kleppner and R. Jackiw, {\em Science} {\bf 289}, 893 (2000)

\bibitem{Tulin14}  M. Peshkin and A. Tonomura, {\em The Aharonov-Bohm Effect.} Springer, New York, (1989).

\bibitem{Tulin15} M. Peshkin, {\em Foun.  Phys.} {\bf 29}, 481 (1999).  

\bibitem{Tulin16} T. H. Boyer, {\em Foun.  Phys.} {\bf 30}, 893 (2000); {\em Foun.  Phys.} {\bf 32}, 41 (2002)

\bibitem{Tulin17} A. Tonomura and F. Nori, {\em Nature} {\bf 452} 298 (2008). 

\bibitem{Tulin18} A. Caprez, B. Barwick, and H. Batelaan, {\em Phys. Rev. Lett.} {\bf 99}, 210401 (2007).

\bibitem{Tulin19} B. Reulet, M. Ramin, H. Bouchiat, and D. Mailly, {\em Phys. Rev. Lett.} {\bf 75}, 124 (1995); R. Deblock et al., {\em Phys. Rev. Lett.} {\bf 89}, 206803 (2002). 

\bibitem{Tulin20} V.L. Gurtovoi et al., {\em in Proceedings of 16th International Symposium "NANOSTRUCTURES: Physics and Technology"} Vladivostok, Institute of Automation and Control Processes, 2008, p.247. 

\bibitem{Tulin21} S.V. Dubonos et al., {\em in Proceedings of 15th International Symposium "NANOSTRUCTURES: Physics and Technology"} St Petersburg: Ioffe Institute, 2007 p. 60. 

\bibitem{Tulin22} I.~O.~Kulik, {\em Zh.Eksp.Teor.Fiz.} {\bf 58}, 2171 (1970); {\em Pisma Zh.Eksp.Teor.Fiz.} {\bf 11}, 407 (1970) ({\em JETP Lett.} {\bf 11}, 275 (1970)).

\bibitem{Tulin23} F.~von Oppen and E.~K.~Riedel, {\em Phys. Rev.Lett.} {\bf 66}, 587 (1991).

\bibitem{Tulin24} F.~von Oppen and E.~K.~Riedel, {\em Phys. Rev. B} {\bf 46}, 3203 (1992). 

\bibitem{Tulin25} A.~A.~Burlakov et al.,  {\em Pisma Zh.Eksp.Teor.Fiz.} {\bf 86}, 589 (2007) ({\em JETP Lett.} {\bf 86}, 517 (2007)).

\bibitem{ Tulin26} M.~Tinkham, {\em Introduction to Superconductivity.} McGraw-Hill Book Company (1975).

\end{thebibliography}
\end{document}